\documentclass[sigconf]{acmart}
\usepackage{color}
\usepackage{multirow}
\usepackage{graphicx}
\usepackage{amsmath}

\AtBeginDocument{%
  \providecommand\BibTeX{{%
    \normalfont B\kern-0.5em{\scshape i\kern-0.25em b}\kern-0.8em\TeX}}}

\newcommand{\etal}{\textit{et al.}}
\setcopyright{none}
\begin{document}

\title{MIC: Model-agnostic Integrated Cross-channel Recommender}



\settopmatter{authorsperrow=1} 
\newcommand{\tsc}[1]{\textsuperscript{#1}} 
\author{Yujie Lu\tsc{*, 1}, Ping Nie\tsc{*, 3}, Shengyu Zhang\tsc{2}, Ming Zhao\tsc{3}, Ruobing Xie\tsc{3}, William Wang\tsc{1}, Yi Ren\tsc{3}}
\affiliation{
  \institution{\tsc{1}University of California, Santa Barbara \country{USA}} 
  \institution{\tsc{2}Zhejiang University \country{China}}
  \institution{\tsc{3}Tencent \country{China}}
}

  




\renewcommand{\shortauthors}{Trovato and Tobin, et al.}
\newcommand{\methodname}[1]{MIC}
\begin{abstract}
Semantically connecting users and items is a fundamental problem for the matching stage of an industrial recommender system.
Recent advances in this topic are based on multi-channel retrieval to efficiently measure users' interest on items from the massive candidate pool.
However, existing studies are primarily built upon pre-defined retrieval channels, including User-CF (U2U), Item-CF (I2I), and Embedding-based Retrieval (U2I), thus access to the limited correlation between users and items which solely entail from partial information of latent interactions.
In this paper, we propose a model-agnostic integrated cross-channel (\methodname~) approach for the large-scale recommendation, which maximally leverages the inherent multi-channel mutual information to enhance the matching performance.
Specifically, \methodname~ robustly models correlation within user-item, user-user, and item-item from latent interactions in a universal schema.
For each channel, \methodname~ naturally aligns pairs with semantic similarity and distinguishes them otherwise with more uniform anisotropic representation space.
While state-of-the-art methods require specific architectural design, \methodname~ intuitively considers them as a whole by enabling the complete information flow among users and items.
Thus \methodname~ can be easily plugged into other retrieval recommender systems.
Extensive experiments show that our \methodname~ helps several state-of-the-art models boost their performance on four real-world benchmarks.
The satisfactory deployment of the proposed \methodname~ on industrial online services empirically proves its scalability and flexibility.
\end{abstract}




\maketitle
\section{Introduction}

\begin{figure}[h]
    \centering
    \includegraphics[width=\linewidth]{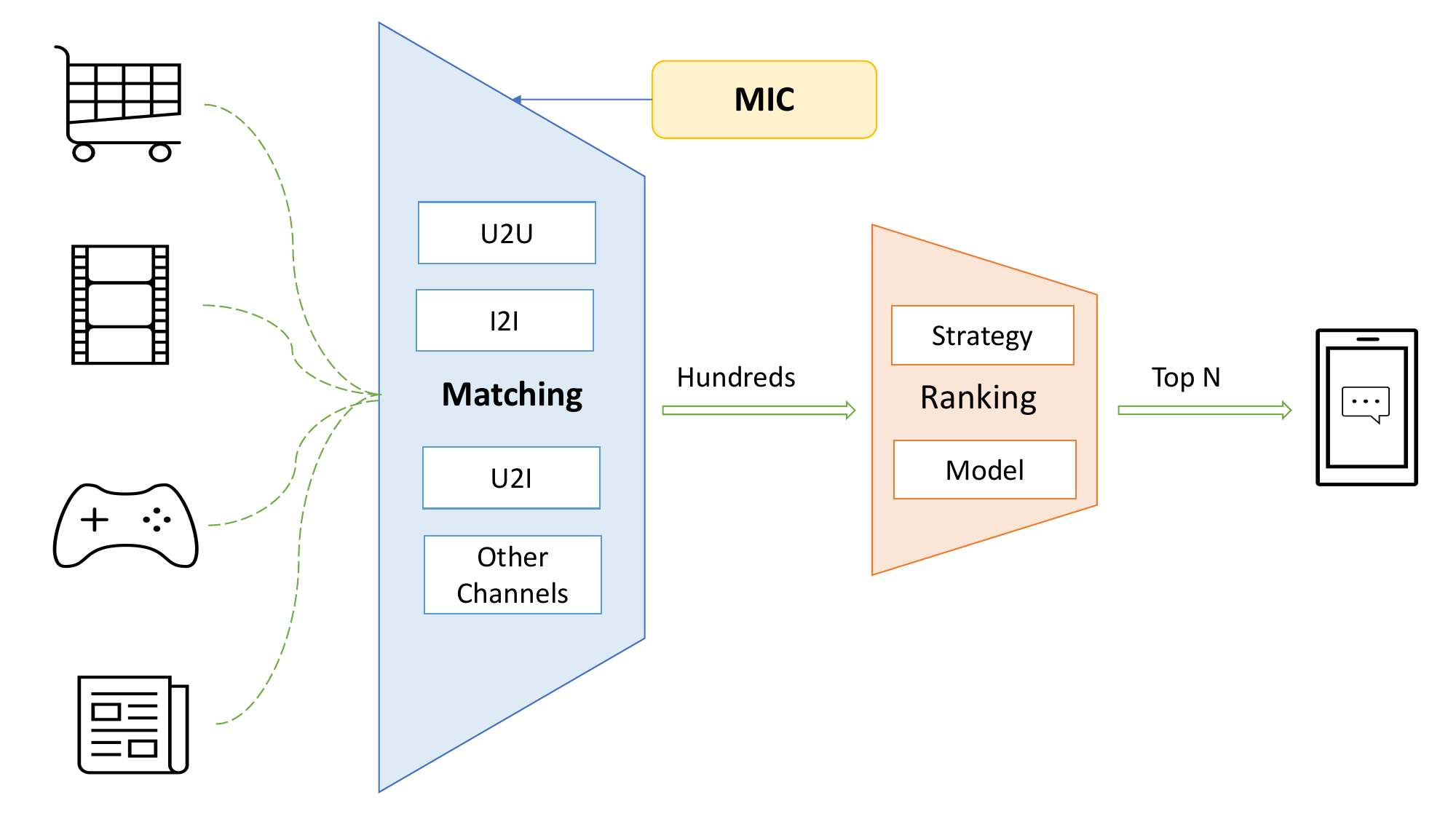}
    \caption{A diagram of a typical two-stage (matching and ranking) recommender system in the real world. \methodname~ can be easily applied in the matching stage.
    }
    \label{fig:recommender}
\end{figure}

\begin{figure}[t]
    \centering
    \includegraphics[width=\linewidth]{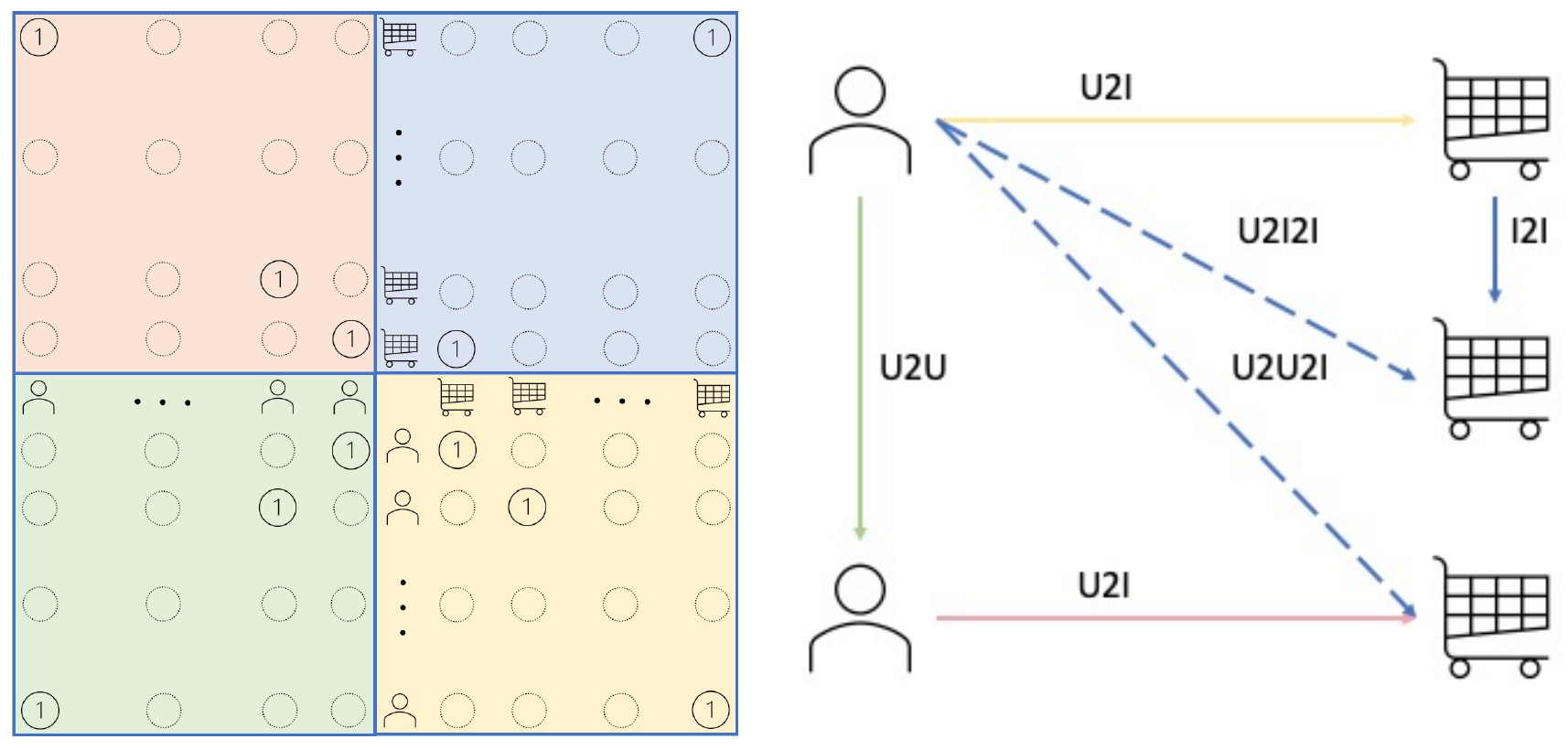}
    \caption{A diagram for multiple connected paths (U2I, I2I, U2U, U2U2I, U2I2I) among users and items. The interactions and correlations are reflected in the left matrix.
    } 
    \label{fig:crosschannel}
\end{figure}

In this era of information explosion, recommendation services have emerged to match various products with diverse users efficiently.
As shown in Figure~\ref{fig:recommender},
the matching stage providing the retrieved items list to the ranking stage is the cornerstone and the bottleneck of a typical two-stage industrial recommender system.
Figure~\ref{fig:crosschannel} depicts the commonly used retrieval connected paths: \textit{1)} U2I: Directly recommend items to users. \textit{2)} I2I: Recommend similar items.
\textit{3)} U2U: Retrieve similar users. \textit{4)} U2U2I: Recommend items that similar users like based on user-based collaborative filtering. \textit{5)} U2I2I: Recommend similar items based on user interaction history and similar items.
These paths finally depict the commonly retrieval channels: \textbf{U2U} (by U2U2I path), \textbf{U2I} (by U2I path) and \textbf{I2I} (by U2I2I path).
In this scenario, it is vital to efficiently model user preferences over items to retrieve from large-scale candidate pools; thus, multi-channel retrieval, which efficiently mixes the diversified retrieved items, is a natural and indispensable approach.

However, most previous methods seek to improve the performance of user modeling based on a single channel, thus failing to leverage inherent correlations in the user-based channel, item-based channel, and user-item channel simultaneously.
For user channel (U2U), it is common in industry recommendation system to use Locality sensitive hashing \cite{gionis1999lsh}, Paragraph2Vector \cite{pmlr-v32-le14} and DSSM \cite{huang2013learning} models to encode user history items and generate similar users.
\cite{Li2021PathbasedDN} improve the performance of personalization and diversity in item-based collaborative filtering from the item channel (I2I) perspective.
\cite{paul2016youtubednn,Huang2013LearningDS, li2019mind,cen2020cmind,Lu2021FutureAwareDT} are proposed to model dynamic and diversified user preferences based on interactions records from the user-item channel (U2I).
For retrieval from multiple sources, \cite{ruobing2021HRL} propose a hierarchical reinforcement learning framework to recommend heterogeneous items.
Nevertheless, the existing method focuses on improving performance based on partial information from each channel, significantly reducing their performance.


We argue that addressing the aforementioned issues in a unified manner is under-explored and points to a new promising direction for developing recommender systems.
Models that solely focus on a single angle could learn common relevance between users and items while ignoring the inherent cross-channel information and performing poorly in a real-world scenario.
Industrial systems attempt to mitigate such performance reduction by retrieving items based on multiple channels, including various features, strategies, and models.
However, existing offline training pipelines are bound to a channel-specific model framework, and the online mixture of multiple channels retrieval is usually controlled by a simple quota mechanism, which leads to two major challenges: \textit{a)} Devising a mechanism to utilize cross-channel information. \textit{b)} Improving item retrieval accuracy and diversity simultaneously in a unified manner.
In contrast, our proposed model-agnostic integrated cross-channel (\methodname~) approach is towards addressing the challenges mentioned above within a universal retrieval recommender system.


In this work, we focus on capturing correlations among users and items across multiple channels with a single model in a unified schema. To achieve this, we first found that it is possible to use one model such as Comirec \cite{cen2020cmind} for three-channel retrieval: U2I, U2U, I2I. Then we designed cross-channel contrastive learning techniques to boost a single model's performance on three channels.
We introduce cross-channel contrastive learning techniques into our unified framework with learnable and configurable settings to handle the dynamic and uncertain nature when connecting users and items. 
In particular, we randomly perturb the fields of each instance and perform dropout in the embedded feature space.
The objective is to learn the representations by leveraging a contrastive learning loss to maximize the similarity between the embeddings of two versions of the same instance.
User and item representations are learned in their own semantic space via \textbf{intra-channel} contrastive loss with the user-user (U-U) contrastive and the item-item (I-I) contrastive training setting.
To further connect users and items, we intuitively perform a non-linear projection to learn additional users and items representations in a common semantic space via \textbf{inter-channel} user-item (U-I) contrastive loss. The relevance between users and items is measured as the cosine similarity between their vectors in a shared space. Finally, We built a unified score function to generate top-$N$ items from U2I, U2U, I2I retrieved items.

\begin{figure*}[htb]
    \centering
    \includegraphics[width=\linewidth]{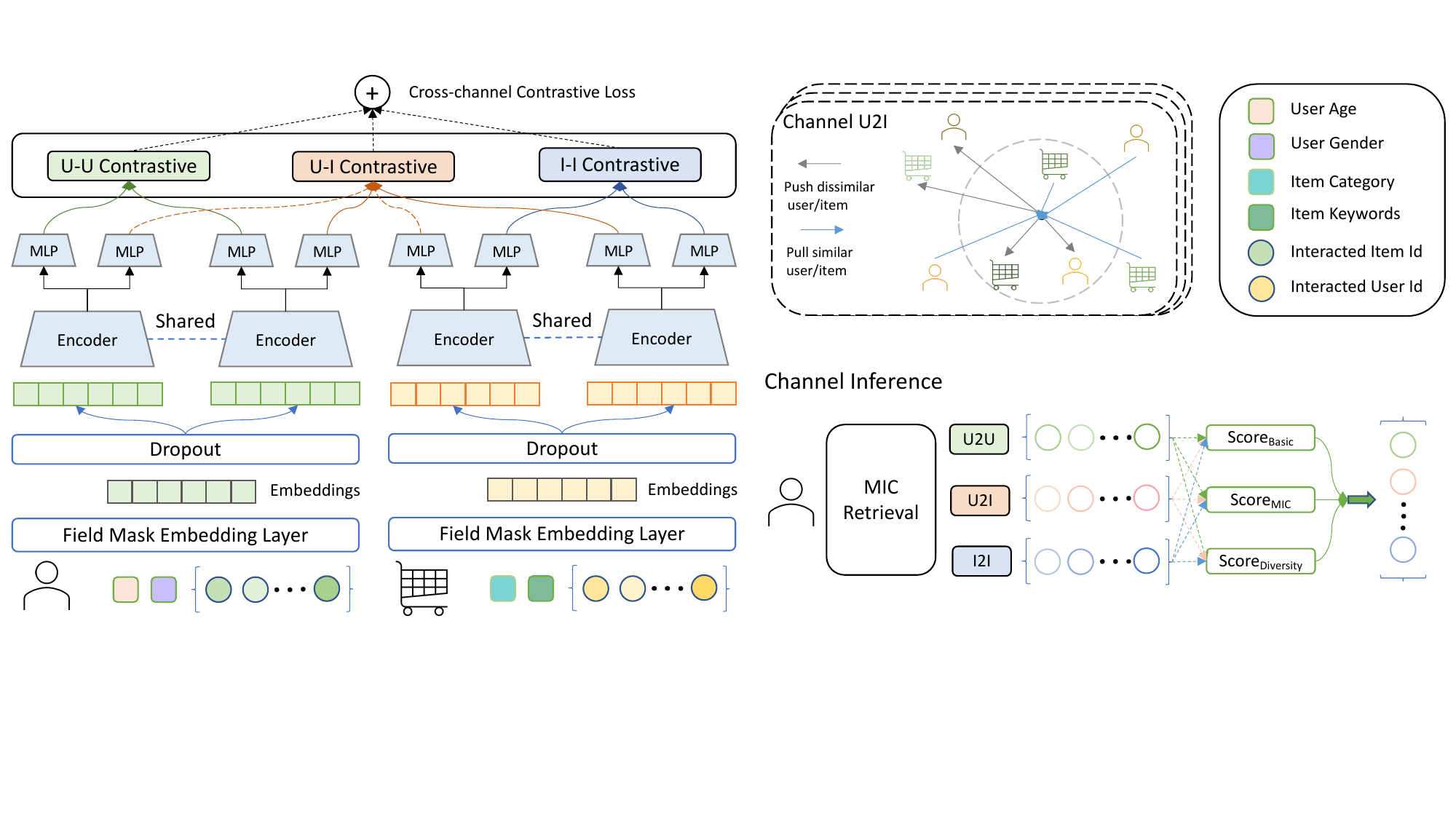}
    \caption{Overview of model-agnostic integrated cross-channel recommenders (\methodname~). The perturbations is performed in both field level and embeded features level. The user-item (U2I), user-user (U2U) and item-item (I2I) modules are aggregated to calculate cross-channel contrastive loss.
    In Inference stage, \methodname~ applies aggregation over items retrieved from three channels and compute Score$_{Basic}$, Score$_{MIC}$ and Score$_{Diversity}$ for final recommendation reference.
    } 
    \label{fig:ICLR}
\end{figure*}

\methodname~ can realize efficient multi-channel retrieval to capture the co-evolving diversified and dynamic users and items representations in an integrated schema.
Since the cross-channel learning module is independent of the encoders and the embedding layer is adaptable to sparse and dense features of users and items, \methodname~ achieves a model-agnostic performance boost by simply switching the encoder to other retrieval models as shown in Figure~\ref{fig:ICLR}. To summarize, the main contributions of this work are as follows:

\begin{itemize}
    \item We formulate the matching stage of recommendation as connecting user and item from multiple channels and propose a model-agnostic \methodname~ architecture based on integrated cross-channel user and item representation learning techniques. 
    \item We address the aforementioned long-standing challenges in recommendation in a unified manner via a cross-channel contrastive aggregation mechanism.  \methodname~ mitigates the uncertainty of co-evolving user-item correlations and alleviates the seesaw effect between retrieval accuracy and diversity. To the best of our knowledge, this is the first work that proves it is possible to simultaneously utilize U2I, U2U, and U2I channels to improve retrieval accuracy and diversity.
    \item Compared with the existing method, \methodname~ shows superior effectiveness and efficiency performance on four public datasets. \methodname~ can also be incorporated into other matching stage recommenders to boost their performance.
    \item We deployed \methodname~ on the Tencent News platform, the satisfactory online $A/B$ test results on million-scale users and items confirm the efficiency and effectiveness of \methodname~ practiclly.
\end{itemize}


\section{Approach}
\subsection{Problem Formulation}
In a typical recommendation scenario, we have a set of users and a set of items which can be denoted as ${U=\{u_1, u_2, ..., u_{|U|}\}}$ and ${V=\{v_1, v_2, ..., v_{|V|}\}}$, respectively. Let ${X_u = \{ x_1^u, x_2^u, ..., x_{|X_u|}^u \}}$ denote the sequence of interacted items from user ${u \in U}$ sorted in a chronological order: ${x_t^u}$ denotes the item that the user ${u}$ has interacted with item at time step ${t}$. Given the user historical behaviors, the goal of the sequential recommendation task considered in this paper is to retrieve a subset of items from the pool ${V}$ for each user in ${U}$ such that the user is most likely to interact with the recommended items.
Specifically, each instance is represented by a tuple ${(X_u, F_u, F_v)}$, where ${X_u}$ denotes the interactions records of user ${u}$, ${F_u}$ denotes the fields of features of the user ${u}$ including user ID, gender and age. ${F_v}$ denotes the fields of features of target item ${v}$ including the information of item ID, item keywords.
\methodname~ learns a function ${f}$ and ${g}$ for the representations of users and items respectively as
\begin{equation}
    \overrightarrow{e_u} = f(X_u, F_u), \overrightarrow{e_v} = g(F_v)
\label{eq:userRepFunc}
\end{equation}
where ${\overrightarrow{e_u} \in \mathbb{R}^{d\times1}}$ denotes the representation vector of user $u$, and $d$ is the dimension. ${\overrightarrow{e_v} \in \mathbb{R}^{d\times1}}$ denotes the representation vector of item $v$.
When user representation vector and item representation vector are learned, top-N items are recommended according to the likelihood function ${p}$ as:

\begin{equation}
    p(i|U, V, X) = \lambda_{u2v} * p(\overrightarrow{e_u}, \overrightarrow{e_v}) + \lambda_{u2u} * p(\overrightarrow{e_u}, U, X) + \lambda_{v2v} * p(\overrightarrow{e_v}, X)
\end{equation}

where $N$ is the predefined number of items to be retrieved. ${\overrightarrow{e_v}}$ is the embedding of item v from a set of items $V$. $\lambda_{u2v}$, $\lambda_{u2u}$ and $\lambda_{v2v}$ represent the balance factor for each inference channel U2I, U2U and I2I respectively.
We use Grid Search to choose these hyperparameters.
As we mainly focus on improving the performance in the matching stage of classical industrial recommender systems,
Our framework outputs the probabilities for all the items, representing how likely the specific user will engage with these items, and retrieves top-N candidate items.

\subsection{Datastore and Inference Procedure}
When the MIC is trained, we can predict all users' and items' representation in the training dataset and build a user Datastore and an item Datastore. In the user Datastore, we define the key-value pair $(\overrightarrow{e_u}, u)$ where the key $\overrightarrow{e_u}$ is the vector representation of the value user $u$. In the item Datastore, the key-value pair is $(\overrightarrow{e_v}, v)$ from the item $v$ representation $\overrightarrow{e_v}$. We also build an interaction Datastore with key-value pairs $(u, X_u)$  where the key is the user ID, and the value is the user interaction history. 

At test time, given the user $u$ with interaction history and features, we get user representations $\overrightarrow{e_u}$ from $f(X_u, F_u)$. MIC uses $e_u$ to retrieve $N$ items from item Datastore (U2I) and $m$ similar users from user Datastore. For each similar user, we obtain their interaction history from the interaction Datastore (U2U). We also search similar items according to the user's history from item Datastore (I2I). After U2I, U2U, and I2I channels' search, we have a set of candidate items with counting scores $V_C = \{(v_i, s_i)\}$, where $s_i$ is the retrieved items' repeated number. If a specific item is retrieved from more similar users or more similar user interactions, then the counting score will be larger. The counting scores directly considers the contribution of U2U and I2I channel. The size of $V_C$ is often larger than N and much smaller than $|V|$. MIC calculates each item's probability with user embedding, item embedding, and item counting numbers. 

\begin{equation}
    Score_{Basic}(v_i) = p(\overrightarrow{e_{v_i}, u}), Score_{MIC}(v_i) = \frac{exp(s_i)}{\sum_{j \in |V_C|}{exp(s_j)}}
\end{equation}

\begin{equation}
    g(i,j) = \delta(C(i) != C(j),
    \\ Score_{Div}(v_i) = \sum_{i\in V_C} \sum_{j \in V_C} g(i,j)
    \label{eq:dscore}
\end{equation}

\begin{equation}
    Score = Score_{Basic} + \lambda_{mic} Score_{MIC}(v_i) + \lambda_{div} Score_{Div}(v_i)
    \label{eq:finalscore}
\end{equation}

where $\lambda_{mic}$ represents the adjustable factor to aggregate items from different channels and $\lambda_{diversity}$ to control retrieved items' diversity.
Similar to ComiRec~\citep{cen2020cmind}, we control retrieved items' diversity according to item category.
We use Grid Search to choose $\lambda_{mic}$ and $\lambda_{diversity}$.
$C$ denotes the category of the specific item.
After MIC scored each item to the current user according to U2I, U2U, and I2I channels results, we choose top N items from $V_C$. 

\subsection{Overall Architecture}
Figure~\ref{fig:ICLR} gives an overview of our proposed \methodname~ model in each component. \methodname~ is composed of 1) Perturbation Mining module: Perturbing data samples via Dropout Layer and Field Mask Embedding Layer, and retrieving similar samples via Nearest Neighbor Mining to construct contrastive positive pairs. 2) Encoder Module: Encoding the user and item features into inherent representations;  Replaceable with existing encoders from retrieval baselines. 3) Cross-channel Contrastive module: Maximally leveraging the inherent mutual information in multiple channels via contrastive loss from user-user, item-item, and user-item space.
In each channel module, the objective is to pull similar samples and push away dissimilar ones.

\subsection{Perturbating and Mining}
Contrastive learning method encourages positive pairs to have similar representations while negative pairs to have dissimilar representations. In the scenario of our unified framework, we consider both users and items as the anchor and generate pseudo views of each instance for comparison. We also leverage retrieved nearest neighbors to support the augmented sample views further.

\subsubsection{Multi-level Perturbation}

Data augmentation has been proved effective and widely used in contrastive prediction tasks without changing the architecture \cite{chen2020simple}.
We devise a simple augmentation method to decouple from the neural network architecture.
For users, we randomly masked the user fields, including attributes (Id, gender, age) and interaction sequence (item Id).
Similarly, we randomly masked attributes (item Id, keywords) and each item's interaction records (user Id).
In addition to the field-level perturbations, the dropout is performed in the embedded features space.
When only perturbation-based view augmentation is available, we treat the other ${2(N-1)}$ augmented examples within a minibatch as negative examples.

\subsubsection{Nearest Neighbor Mining}
We observe limited views generated by augmentation.
First, view augmentation is limited to origin instance and fail to provide diversified samples.
Second, effective augmentation is difficult to devise, refine, and evaluate in some scenarios.
Finally, the augmentation method suffers from the balance between providing diversified views and maintaining semantic consistency.

In addition to augmentation, we argue that it's necessary to leverage information from a retrieval angle of view.
For users, we retrieve the anchor user's k-nearest neighbor (kNN) in the representation space as the extension of user positive pairs.
Besides, we adopt k-means++ to cluster the users and choose users from different clusters as hard negative samples.
For items, both positive and hard negative samples are mined in the representation space in the same manner as users.
At the interaction level, we use users to retrieve items and items to retrieve users. Before that, we project user and item representation in the same space. The same retrieval is then applied in this joint user-item representation space.
Note that our sample selection pool is highly flexible. All the parameters, including the number of nearest-neighbor, number of clusters, and number of masked attributes, are tuned during training and adaptable to manual modification.
Thus \methodname~ maintains scalability and robust temporal efficacy in fast-speed transforming online changes.

\subsection{Cross-channel Contrastive Estimation}
Many works~\cite{He2017NeuralCF} directly optimize by forcing $click(u,v) = 1$ in diagonal and $click(u,v) = 0$ in other positions. However, these forcing methods assume the deterministic correlation between user and items, which is always not true in the real world. 
The real-world environment is always stochastic (\textit{e.g.} diversified and dynamic user behaviors), where deterministic functions can only predict the average. 
On the other hand, contrastive estimation is an energy-based model. Instead of setting the cost function to be zero only when the prediction and the observation are the same, the energy-based model assigns low cost to all compatible prediction-observation pairs. Thus, the contrastive estimation can handle the stochasticity by its nature~\cite{lecun2006tutorial}. Inspired by recent contrastive learning algorithms~\cite{chen2020simple}, we propose to train these models by maximizing agreement between the anchor and augmented views via a contrastive loss. We randomly sample a minibatch of $N$ user-item pairs $(u, i)$.
For the unified model, augmented users and items and the mined samples in the support set are defined as positive examples.
Following SimCLR~\cite{chen2020simple}, we treat the other $2(N - 1)$ real representation within a minibatch as negative examples. We use cosine similarity to denote the distance between two representation $(u, v)$, that is $\texttt{sim}(u,v) = \mathbf{u}^T \cdot \mathbf{v}/||\mathbf{u}|| \cdot ||\mathbf{v}||$. The loss function for a positive pair of examples $(u, v)$ is defined as:
\begin{equation}
\mathcal{L}_{uv} = -\text{log} \frac{\text{exp}(\texttt{sim}(u, v_i)/\tau)}{\sum_{\substack{j=1 \\ j\neq i}}^{N} \text{exp}(\texttt{sim}(u, \tilde{v_j})/\tau)} -\text{log} \frac{\text{exp}(\texttt{sim}(v, u_i)/\tau)}{\sum_{\substack{j=1 \\ j\neq i}}^{N} \text{exp}(\texttt{sim}(v, \tilde{u_j})/\tau)}
\label{eq:useritem_contrastive}
\end{equation}
where $\tau$ denotes a temperature parameter that is empirically chosen as $0.1$.

Similarly, for user-user and item-item model, the loss function for a positive pair of examples $(\tilde{u}, u)$ and $(\tilde{v}, v)$ is defined as:

\begin{equation}
\mathcal{L}_{uu} = -\text{log} \frac{\text{exp}(\texttt{sim}(u_k, \tilde{u_k})/\tau)}{\sum_{\substack{j=1 \\ j\neq k}}^{N} \text{exp}(\texttt{sim}(u_k, u_j)/\tau)}
\label{eq:item_contrastive}
\end{equation} 

\begin{equation}
\mathcal{L}_{vv} = -\text{log} \frac{\text{exp}(\texttt{sim}(v, \tilde{v_i})/\tau)}{\sum_{\substack{j=1 \\ j\neq i}}^{N} \text{exp}(\texttt{sim}(v, v_j)/\tau)}
\label{eq:user_contrastive}
\end{equation} 

The basic logistic loss by comparing the cosine similarity of users and items are computed as below:
\begin{equation}
\mathcal{L}_{basic} = -\frac{1}{N}\sum_{i} \left[ y_{i} \log \hat{y_i} + (1-y_{i}\log(1-\hat{y_i})) \right]
\end{equation}

\subsection{Integrated Model}
The user-item (U2I), user-user (U2U) and item-item (I2I) modules are aggregated to calculate cross-channel contrastive loss. We use the Adam optimizer to train our method. The objective function for training our model is to minimize the following cross-channel contrastive loss:

\begin{equation}
\mathcal{L} = \lambda \mathcal{L}_{basic} + (1-\lambda)(\mathcal{L}_{uv} + \mathcal{L}_{vv} + \mathcal{L}_{uu})
\label{eq:total_loss}
\end{equation}
where $\lambda$ is set to $0.7$, each channel weight is $1:1:1$ after parameter optimization in our experiments. \methodname~ can achieve the optimum trade-off across multiple channels by selecting the value of hyperparameter $\lambda$ and channel weight. During training, the total loss is computed across all positive pairs in a mini-batch.

\subsection{Model-agnostic Plugin}

\methodname~ can also be treated as a plug-in to other matching stage recommenders by simply switching the encoder.
\methodname~ incorporate the perturbation and mining module in the item-side and add a cross-channel contrastive learning module on top of the retrieval baselines.
Since the cross-channel learning module is independent of the encoders and the embedding layer is adaptable to sparse and dense features of users and items, \methodname~ is highly flexible and achieves a model-agnostic performance boost in retrieving items from multiple channels efficiently.

\subsection{Cross-channel Inference}

During the inference phase of \methodname~, we get user and item representation from the user and item side encoder, respectively.
For the U2I channel, we directly use the user vector to retrieve $K_1$ nearest neighbor from the whole item pool.
For the U2U channel, we search $M_1$ similar users from the training dataset and retrieve $K_2$ items from $M_1$ similar users' history by considering the weight of similar users and user-item vector cosine similarity.
For the I2I channel, we use the user's history to find $M_2$ relevant items within the whole item vector space for each history item and retrieve $K_3$ items by considering the weight of similar items and user-item vector cosine similarity.
Finally, according to the final score in Equation~\ref{eq:finalscore}, we rank top $N$ items from multiple channels ($K_1 + K_2 + K_3$).

\subsection{Online Deployment} \label{online deployment}

We have deployed MIC on a well-known platform named Tencent News. Tencent News is one of the most popular news recommendation software, which has more than 300 million active users per month. The online architecture of Tencent News mainly consists of the retrieval stage and ranking stage widely used in the industry. The retrieval stage aims to quickly search hundreds of candidates from the entire news corpus (containing million-level news) efficiently, while the ranking stage aims to score news items accurately.
MIC is deployed on the retrieval stage and an embedding-based recall model. We train and update our MIC model hourly.

Once MIC is trained, we infer all item vectors in the corpus and users' vectors of the current hour. Item vectors and user vectors are used to search similar items and users offline. Similarities of each item and user, user's interaction history are stored in Redis \footnote{\href{https://redis.io/}{https://redis.io/}}. Item vectors are also used to build the item Faiss \footnote{\href{https://github.com/facebookresearch/faiss}{https://github.com/facebookresearch/faiss}} search index. MIC is also served online for real-time user representation generating. When a user request comes, MIC first builds Redis key with userID and last $M$ interaction items (we keep $M_3$ similar items for each item), then get $M_1$ similar users (we keep last $M_2$ clicked items for each user). For each similar user, MIC gets their clicked items from Redis. So we get $K_2 = M_1 \times M_2$ items from similar users (U2U) and $K_3 = M \times M_3$ items from similar items (I2I). We package real-time user features and generate user representation from MIC online serving then the representation is used to search top $K_1$ items from item Faiss index. Finally, the $K_1 + K_2 + K_3$ items are aggregated according to final score in Equation~\ref{eq:finalscore} and top $N$ items are recalled. 
The fast nearest neighbor retrieval of Faiss search time is T1 (less than ten milliseconds). The time cost of similar users and items from Redis is T2 ( less than ten milliseconds). The aggregation step time T3 (less than ten milliseconds). The whole time cost is acceptable for online serving in our system. 

While our MIC model is updated hourly, we are still able to capture the real-time user preferences. We build real-time MIC request features including real-time click history for each user representation. User and corresponding similar users' interactions stored in Redis are also updated (in seconds) in real-time.
So we can use real-time clicked items to  get similar items from I2I channel and similar users' real-time interactions from U2U channel.
Since the time cost is also related to similar users and similar items, the hyper-parameters should be adjusted to satisfy the online serving time requirements. In our system, $M_1=40$, $M_2=60$, $M_3=50$, $M=30$ and $N = 200$. 

\begin{table*}
\centering
\caption{Performance of four public datasets: Amazon Book, Taobao, Movielens and Steam. Results of three retrieval baselines and the proposed \methodname~ are reported over three metrics: Recall, NDCG and Hit Rate. \textbf{Gain} represents the performance gain of $X$+\methodname~ over vanilla $X$ model.}
\label{tab:main_table}
\begin{tabular}{cccccc||cccccc} 
\toprule
\multirow{2}{*}{\begin{tabular}[c]{@{}c@{}}\\Datasets\end{tabular}}    & \multirow{2}{*}{@N}  & \multirow{2}{*}{Metrics} & \multicolumn{3}{c||}{Baselines} & \multicolumn{6}{c}{\textbf{$X$+MIC}}                                                 \\ 
\cmidrule{4-12}
                                                                       &                      &                          & DNN    & Gru4Rec & ComiRec      & DNN    & \textbf{Gain}    & Gru4Rec & \textbf{Gain}    & ComiRec & \textbf{Gain}     \\ 
\hline
\multirow{6}{*}{\begin{tabular}[c]{@{}c@{}}Amazon \\Book\end{tabular}} & \multirow{3}{*}{@20} & Recall                   & 5.608  & 5.877   & 6.634        & 5.934  & \textbf{5.81\%}  & 6.0141  & \textbf{2.33\%}  & 7.457   & \textbf{12.41\%}  \\
                                                                       &                      & NDCG                     & 5.371  & 5.835   & 6.023        & 5.836  & \textbf{8.66\%}  & 5.992   & \textbf{2.69\%}  & 6.195   & \textbf{2.86\%}   \\
                                                                       &                      & Hit Rate                 & 12.291 & 12.545  & 13.423       & 12.828 & \textbf{4.37\%}  & 12.997  & \textbf{3.60\%}  & 15.124  & \textbf{12.67\%}  \\ 
\cline{2-2}
                                                                       & \multirow{3}{*}{@50} & Recall                   & 8.885  & 8.908   & 10.2574      & 9.3066 & \textbf{4.75\%}  & 9.411   & \textbf{5.65\%}  & 11.55   & \textbf{10.90\%}  \\
                                                                       &                      & NDCG                     & 6.594  & 6.915   & 7.217        & 7.077  & \textbf{7.32\%}  & 7.105   & \textbf{2.75\%}  & 7.889   & \textbf{9.31\%}   \\
                                                                       &                      & Hit Rate                 & 18.709 & 18.949  & 19.231       & 19.373 & \textbf{3.55\%}  & 19.535  & \textbf{3.09\%}  & 22.790  & \textbf{18.51\%}  \\ 
\hline
\multirow{6}{*}{Taobao}                                                & \multirow{3}{*}{@20} & Recall                   & 3.319  & 4.132   & 5.065        & 3.531  & \textbf{6.39\%}  & 4.442   & \textbf{7.50\%}  & 5.642   & \textbf{11.39\%}  \\
                                                                       &                      & NDCG                     & 12.493 & 15.449  & 19.324       & 13.481 & \textbf{7.91\%}  & 17.995  & \textbf{16.48\%} & 21.221  & \textbf{9.82\%}   \\
                                                                       &                      & Hit Rate                 & 28.417 & 32.033  & 38.429       & 29.592 & \textbf{4.13\%}  & 36.661  & \textbf{14.45\%} & 41.878  & \textbf{8.97\%}   \\ 
\cline{2-2}
                                                                       & \multirow{3}{*}{@50} & Recall                   & 5.075  & 6.118   & 7.115        & 5.278  & \textbf{4.00\%}  & 6.377   & \textbf{4.23\%}  & 7.861   & \textbf{10.48\%}  \\
                                                                       &                      & NDCG                     & 14.263 & 16.084  & 20.635       & 15.187 & \textbf{6.48\%}  & 18.999  & \textbf{18.12\%} & 22.509  & \textbf{9.08\%}   \\
                                                                       &                      & Hit Rate                 & 39.31  & 42.114  & 48.094       & 40.324 & \textbf{2.58\%}  & 45.551  & \textbf{8.16\%}  & 51.607  & \textbf{7.30\%}   \\ 
\hline
\multirow{6}{*}{Movielens}                                             & \multirow{3}{*}{@20} & Recall                   & 12.251 & 12.993  & 13.001       & 12.508 & \textbf{2.10\%}  & 13.012  & \textbf{0.15\%}  & 13.322  & \textbf{2.47\%}   \\
                                                                       &                      & NDCG                     & 36.249 & 37.033  & 37.207       & 36.898 & \textbf{1.79\%}  & 37.603  & \textbf{1.54\%}  & 38.186  & \textbf{2.63\%}   \\
                                                                       &                      & Hit Rate                 & 71.688 & 72.344  & 73.772       & 73.841 & \textbf{3.00\%}  & 74.308  & \textbf{2.71\%}  & 76.551  & \textbf{3.77\%}   \\ 
\cline{2-2}
                                                                       & \multirow{3}{*}{@50} & Recall                   & 23.028 & 24.447  & 25.043       & 23.875 & \textbf{3.68\%}  & 25.003  & \textbf{2.27\%}  & 25.927  & \textbf{3.53\%}   \\
                                                                       &                      & NDCG                     & 38.756 & 39.888  & 41.099       & 40.003 & \textbf{3.22\%}  & 41.309  & \textbf{3.56\%}  & 42.109  & \textbf{2.46\%}   \\
                                                                       &                      & Hit Rate                 & 87.245 & 89.705  & 90.138       & 88.907 & \textbf{1.90\%}  & 90.111  & \textbf{0.45\%}  & 91.391  & \textbf{1.39\%}   \\ 
\hline
\multirow{6}{*}{Steam}                                                 & \multirow{3}{*}{@20} & Recall                   & 2.901  & 2.672   & 2.753        & 3.117  & \textbf{7.45\%}  & 2.839   & \textbf{6.25\%}  & 3.009   & \textbf{9.30\%}   \\
                                                                       &                      & NDCG                     & 4.702  & 4.557   & 5.284        & 4.992  & \textbf{6.17\%}  & 5.703   & \textbf{25.15\%} & 5.503   & \textbf{4.14\%}   \\
                                                                       &                      & Hit Rate                 & 10.308 & 9.928   & 11.044       & 10.554 & \textbf{2.39\%}  & 10.422  & \textbf{4.98\%}  & 11.333  & \textbf{2.62\%}   \\ 
\cline{2-2}
                                                                       & \multirow{3}{*}{@50} & Recall                   & 3.671  & 4.432   & 5.021        & 4.288  & \textbf{16.81\%} & 4.775   & \textbf{7.74\%}  & 5.123   & \textbf{2.03\%}   \\
                                                                       &                      & NDCG                     & 5.077  & 4.997   & 6.23         & 5.779  & \textbf{13.83\%} & 5.413   & \textbf{8.32\%}  & 6.671   & \textbf{7.08\%}   \\
                                                                       &                      & Hit Rate                 & 12.031 & 11.089  & 13.149       & 12.608 & \textbf{4.80\%}  & 12.307  & \textbf{10.98\%} & 14.388  & \textbf{9.42\%}   \\
\bottomrule
\end{tabular}
\end{table*}

\section{Experiments}
\label{sec:EXP}
In this section, we first cover the experimental settings of the dataset, evaluation metrics, parameter settings, and competitors. Then we report the results of extensive offline and online experiments with in-depth analysis to verify the effectiveness of \methodname~.
We conduct experiments to investigate the following research questions:
\begin{itemize}
    \item Research Question 1 (RQ1): How does MIC perform on large public recommendation datasets (Book, Taobao, Movielens, Steam)?
    \item Research Question 2 (RQ2): How does MIC perform in real-word News Recommendations System?
    \item Research Question 3 (RQ3): Are different components and losses essential in MIC?
    \item Research Question 4 (RQ4): How does MIC alleviate the seesaw phenomenon between retrieval accuracy and diversity: Can MIC achieve high retrieval accuracy and diversity simultaneously?
    \item Research Question 5 (RQ5): How does contrastive learning modules (UU,UI,II) help improve the embedding space and recall performance for corresponding U2U, U2I, I2I channel?
\end{itemize}

\subsection{Dataset and Metric}
We used four large benchmark datasets, Amazon Book, Taobao, Movielens, and Steam. The statistics are shown in \ref{tab:staData}.
To compare the performance of different models, we use three metrics \textbf{Recall@N}, \textbf{NDCG@N}(Normalized Discounted Cumulative Gain), and \textbf{HR@N}, where N is set to 20, 50 respectively as metrics for evaluation. In all these three metrics, a higher value implies better performance.
Besides, we adopt a per-user average for each metric.
More details about dataset and evaluation metrics are described in Section~\ref{sec:datasetmetric}.
We track Recall, NDCG, Hit Rate of the Development split during training. Then we keep models with the best Recall Rate on Development split during experiments for a fair comparison.

\begin{table*}[h]
\centering
\caption{Online A/B Test Results. We report the relative performance gain of \methodname~ over Baseline in online A/B experiments.}
\begin{tabular}{cccccll}
\toprule
\#\textbf{Scenario}   & \textbf{EPV ratio} & \textbf{Average Play Percentage} $\uparrow$ &  \textbf{Average Duration} $\uparrow$ & \textbf{Average Viewed Video} $\uparrow$\\
\midrule
Video Recommendation     & 25.00\%                   & +3.51\%                & +1.26\%  &       +1.85\% \\ 
\bottomrule
\end{tabular}
\label{tab:online_results}
\end{table*}

\begin{table}
\caption{Ablation Performance of \methodname~ Variants over ComiRec on Amazon Book dataset with Metric@50.}
\centering
\resizebox{\columnwidth}{!}{%
\begin{tabular}{clcccc} 
\toprule
Modules                                                                        & Settings     & Recall & NDCG  & Hit Rate & Diverstiy  \\
                                                                               & Full Model   & 11.554 & 7.889 & 22.790   & 49.511     \\ 
\hline
\multirow{4}{*}{\begin{tabular}[c]{@{}c@{}}Contrastive\\Loss\end{tabular}} & -UU         & 10.556 & 7.689 & 21.132   & 44.021      \\
                                                                               & -UI         & 10.347 & 6.462 & 21.273   & 42.483      \\
                                                                               & -II         & 11.096 & 7.089 & 22.668   & 46.796      \\
                                                                               & -Perturbation & 8.415  & 5.346 & 16.590    & 34.188     \\
                                                                               & -Mining & 10.176  & 6.098 & 20.727    & 41.983     \\
\hline
\multirow{3}{*}{\begin{tabular}[c]{@{}c@{}}Inference\\Channel\end{tabular}}        & -U2U channel & 11.148 & 7.688 & 22.076   & 45.478     \\
                                                                               & -U2I channel & 11.484 & 7.825 & 22.571   & 45.603     \\
                                                                               & -I2I channel & 11.316 & 7,758 & 22.443   & 41.709     \\
\bottomrule
\end{tabular}
}
\label{tab:ablation}
\end{table}
\subsection{Parameter Settings}
We implement baselines and our proposed model in the same settings for fairness.
The implementation is based on Tensorflow for offline experiments.
The dimension of the collaborative embedding is set as ${128}$. Batch size is set to ${1024}$ on a single NVIDIA P40 GPU. The learning rate is set to ${0.001}$, and the dropout rate is set to ${0.2}$. The temperature parameter is empirically chosen as $0.1$. We utilize Xavier and Adam algorithms in the experiments to initialize and optimize the parameters of the models.

\subsection{Competitors}
\subsubsection{Retrieval Baselines}
YoutubeDNN \cite{paul2016youtubednn} is one of the predominant deep learning models based on collaborative filtering systems incorporating text and image information which have been successfully applied under the industrial scenario.
Gru4Rec~\citep{Hidasi2016SessionbasedRW} is a session-based recommender using Recurrent Neural Networks.
ComiRec \cite{cen2020cmind} is a novel controllable multi-interest framework which can be used in sequential recommendation.

\subsubsection{\methodname~ as Plugin}
As \methodname~ is can also be treated as a model-agnostic plugin, we implement a series of variants with \methodname~ adapted to other retrieval models denoted as $X+\methodname~$.

\subsubsection{\methodname~ Variants}
Our unified model \methodname~ co-learns user and item representation in both shared and their own semantic space. The retrieval model considers mutual information across multiple channels, including use-user, item-item, and user-item channel, simultaneously in an integrated framework.

In addition, we provide three representative variants as \methodname~-UI,\methodname~-UU, and \methodname~-II with single-channel contrastive loss.
For MIC-UI, we add user-item contrastive training on top of ComiRec as a variant of our proposed \methodname~. This variant can capture the information behind the interaction and match the users to appropriate items from the user-item channel.
For MIC-UU, we add user-user contrastive training on top of ComiRec as a variant of our proposed \methodname~. This variant is capable of clustering users and matching similar users to each other from the user channel.
For MIC-II, we add item-item contrastive training on top of ComiRec as a variant of our proposed \methodname~. This variant is capable of clustering items and matching similar items to each other from the item channel.
All compositional ablation results of each contrastive setting are reported in Table~\ref{tab:ablation}.

\subsection{Model-agnostic Gain (RQ1)}
The model performance for the retrieval stage recommender system is shown in Table \ref{tab:main_table}.
We conduct extensive experiments to dissect the effectiveness of our proposed model-agnostic integrated cross-channel (\methodname~) model.
In the baseline performance comparison experiment, the \methodname~ is implemented in a full mode with weighted UI, UU, and II contrastive loss.
All these models are running on the four datasets introduced above: Amazon Book, Taobao, Movielens and Steam.
We plug our \methodname~ into prevalent retrieval baselins: : YouTube DNN, Gru4Rec and ComiRec..
As shown in Table~\ref{tab:main_table}, \methodname~ enhanced models ($X$+MIC) consistently achieve a significant performance gain on all evaluation metrics than the retrieval baselines over four datasets.
In particular, $ComiRec+\methodname~$ gain $10.90\%$, $9.31\%$, $18.51\%$ over vanilla ComiRec model in Recall@50, NDCG@50 and Hit Rate@50 respectively over Amazon Book.

\subsection{Online A/B Test(RQ2)}
We further conduct an online A/B test to evaluate MIC in real-world scenarios. We have deployed MIC on Tencent News Video Recommendation scenarios as stated in Sec \ref{online deployment}. MIC is deployed as a matching model in the retrieval stage, with the remained modules in the whole system unchanged. The online recall baseline is an ensemble model containing tens of retrieval models (embedding-based, rule-based, hot-based, etc.). In the online A/B test, we focus on four metrics, including the exposure page viewed ratio (EPV), Average Play Percentage of each viewed video, Average Duration, and  Average Viewed Video of each user in our platform daily. 
The A/B test was conducted from October 1st, 2021 to October 15th, 2021, and the user number in the experiments group and baseline group is about 1 million. The experimental scenario is Tencent News Video recommendation. 
We report the improvements percentages of MIC in Table \ref{tab:online_results} from which we can know that: 1) MIC achieves significant improvements on Average Viewed Video and Average Duration, which means the recommended videos are more attractive to each user. At the same time, the Average Play Percentage of each video is also improved, which means that MIC provides more precise video to users. 2) The EPV ratio of MIC is about $25\%$, the most effective recall model among all models (the second place recall model's EPV ratio is about $8\%$).

\subsection{Ablation Study (RQ3)}
We conduct ablation experiments of contrastive loss modules and inference channel modules for our proposed \methodname~ enhanced ComiRec~\citep{cen2020cmind}.
Results of variants with various cross-channel contrastive loss settings and various inference channels settings over Amazon Book are reported in Table~\ref{tab:ablation}. 
$-$UU, $-$UI, $-$II represents the Full Model without U-U, U-I, I-I contrastive modules respectively.
$-$Perturbation and $-$Mining represents the Full Model without perturbation and Nearest Neighbor Mining module.
$-$U2U, $-$U2I, $-$I2I represents the Full Model without consideration of retrieved items from U2U, U2I, I2I channel respectively during inference.
We observe performance drop over Recall@50, NDCG@50, HitRate@50 and Diversity in these variants compared with Full Model in Table~\ref{tab:main_table}.
This implies the essential role of each module setting in the Full Model.

\begin{figure}[h]
    \centering
    \includegraphics[width=.95\linewidth]{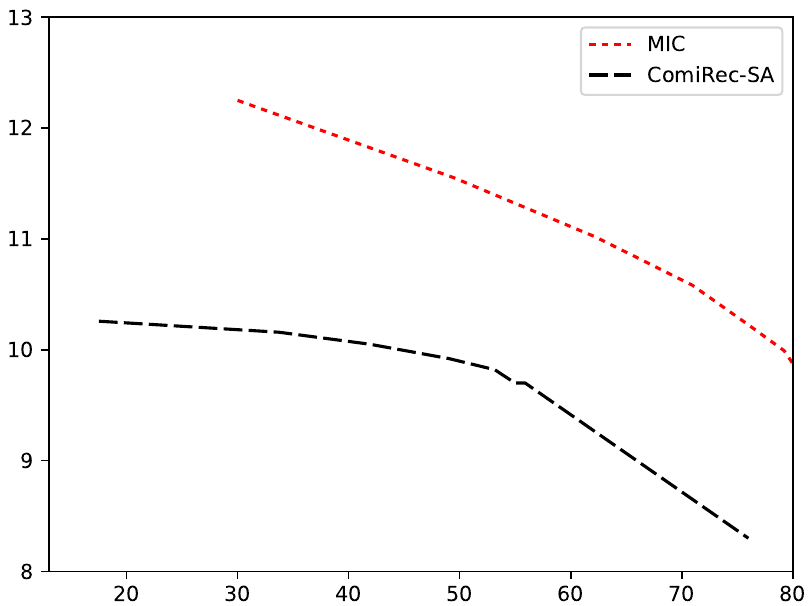}
    \caption{Retrieval Accuracy and Diversity Balance. We compare ComiRec-SA (Black) and MIC enhanced ComiRec-SA (Red) over Amazon Book with Recall@50 (x-axis) and Diversity (y-axis).
    }
    \label{fig:diversity}
\end{figure}

\begin{figure}[ht]
    \centering
    \includegraphics[width=\linewidth]{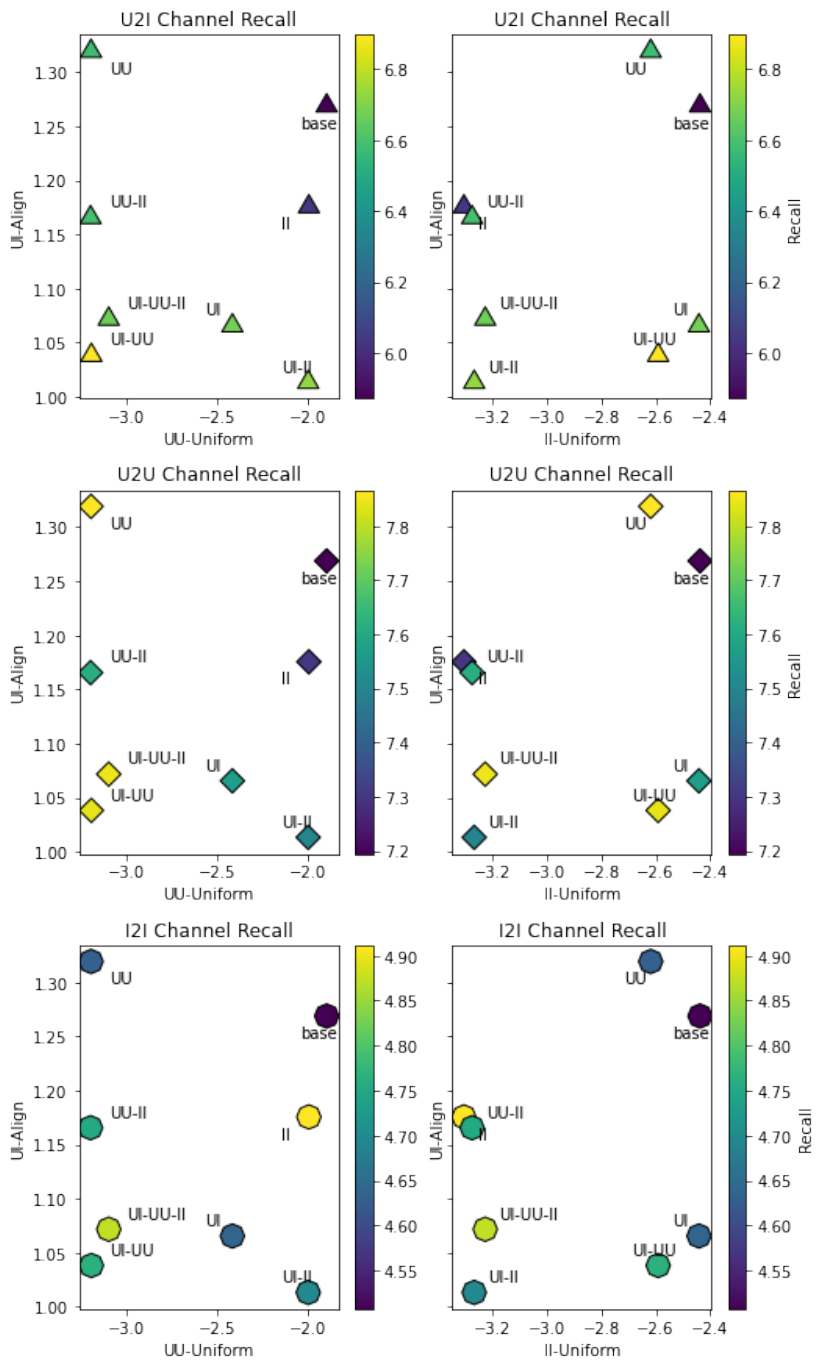}
    \caption{Visualization of User and Item Representation in U2I, U2U and I2I channel over Alignment and Uniformity Metrics of UI-Align, UU-Uniform and II-Uniform and Recall Performance.} 
    \label{fig:align_uniform}
\end{figure}

\subsection{Retrieval Accuracy and Diversity (RQ4)}
There is a Seesaw Effect between retrieval performance and retrieval diversity. We can also observe in Comirec that a better diversity score degrades Recall.
To mitigate this phenomenon, MIC aggregates retrieved items from three channels (U2U, U2I, I2I).
To investigate whether MIC achieve this, we conduct experiments to compare MIC and ComiRec on the Amazon Book dataset.
Results are visualized in Figure~\ref{fig:diversity}.
We can observe that MIC (Red Line) achieves consistent retrieval performance and diversity gain over ComiRec (Black Line).
This indicates that MIC successfully leverages the information to simultaneously improve retrieval performance and diversity.
MIC alleviates the Seesaw Effect and achieves the balance between retrieval accuracy and diversity.

\subsection{Qualitative Results (RQ5)}
While we care about the integrated cross-channel performance of MIC, we still want to see how does contrastive learning
modules (UU,UI,II) help improve the embedding space and
recall performance for corresponding U2U, U2I, I2I channel. We analyze the agreement between user representations, item representations, and final recall performance by the Alignment and Uniformity Metrics \cite{wang2020hypersphere} (lower is better) of UI-Align, UU-Uniform, and II-Uniform.
UI-Align measures the alignment between user and target item representation, UU-Uniform and II-Uniform measure the uniformly distributing of user and item representation, respectively.
As shown in Figure~\ref{fig:align_uniform}, bright yellow denotes better Recall performance.
Each point is marked with corresponding contrastive settings: UI-UU-II means three contrastive learning objects were added, and Base means none contrastive learning objects were considered. For U2I Channel (first row in Figure ~\ref{fig:align_uniform}), the Recall performance is very sensitive to UI-align, and in no doubt, UI-align gets better when UI contrastive learning is considered.
For U2U Channel (second row), UU-Uniform starts to play more important roles besides UI-align. We can find the best recall scores in the bottom left of the "UI-Align, UU-Uniform" graph in U2U Channel Recall. Besides, U2U-Uniform would be better if we added contrastive learning between users.
For I2I Channel (third row), II-Uniform senses to be more important than UI-Align.
The "UI-align, II-Uniform" graph shows that the best Recall appears in the lowest II-Uniform other than the lowest UI-align. 
We observe that if we can simultaneously acquire more aligned user-item representation, and more uniformed user-user, item-item representations, we can push the integrated model's U2I, U2U, and I2I channel performance to the next stage. \methodname~ is one of this type of model-agnostic integrated cross-channel model for recommendations.

\section{Related Works}

\subsection{Recommendation}
Recommendation system can be divided into mainly two categories, content-based recommendation and collaborative filtering.
Collaborative filtering techniques is composed of user-based algorithms \cite{Zhao2010UserBasedCR}, item-based algorithms \cite{Deshpande2004ItembasedTR} and model-based algorithms \cite{Ji2019DeepMS}.
Previous studies \cite{Zheng2016ANA, He2017NeuralCF, Zheng2017JointDM} achieve significant progress based on the idea of user modeling and collaborative recommendation.

Besides collaborative filtering, content-based filtering (e.g. DSSM \cite{Huang2013LearningDS}) is another critical class of recommender systems.
Pure content-based only rely on the feature of users and items, thus ignoring the common preferences shared among similar users and common properties among similar items.
With the emergence of distributed representation learning, user embeddings obtained by neural networks are widely used. \cite{chen2016UPdistributedRep} employs RNN-GRU to learn user embeddings from the temporal ordered review documents. \cite{rakkappan2019stackedrnn} utilizes Stacked Recurrent Neural Networks to capture the evolution of contexts and temporal gaps. \cite{fan2019graph} proposes the framework GraphRec to jointly capture interactions and opinions in the user-item graph.
Due to the intrinsic drawback of both pure content-based and collaborative recommendations, the hybrid model concept is proposed to combine them and benefit each other.
Commonly used hybrid recommendation algorithms include weighted hybrid recommendation algorithm, cross-harmonic recommendation algorithm, and meta-model mixed recommendation algorithm \cite{Bostandjiev2012TasteWeightsAV}.
Dai \etal proposed a dynamic recommendation algorithm \cite{dai2016hybrid} that combines the convolutional neural network and multivariate point process by learning the co-evolutionary model of user-commodity implied features.
Nevertheless, though these hybrid algorithms seek to combine multi-source data, they failed to consider user-user, item-item, and user-item coevolution and relatedness in a unified framework.

\subsection{Contrastive Learning}
Contrastive Learning is a framework to learn representations that obey similarity constraints in a dataset typically organized by similar and dissimilar pairs.
Hadsell et al.~\cite{hadsell2006dimensionality} first proposed to learn representations by contrasting positive pairs against negative pairs.
Some studies ~\cite{wu2018unsupervised, ye2019unsupervised, tian2019contrastive} utilize a memory bank to store the instance class representation vector.
Other work explored the use of in-batch samples for negative sampling instead of a memory bank~\cite{doersch2017multi,ye2019unsupervised, ji2019invariant}
Recently, SimCLR~\cite{chen2020simple} and MoCo~\cite{he2020momentum, chen2020mocov2} achieved state-of-the-art results in self-supervised visual representation learning,  closing the gap with supervised representation learning.
Contrastive training is further explored in
visual representation learning \cite{Yuan2021MultimodalCT, wang2020cloud, Radford2021LearningTV} and views mining~\citep{Azabou2021MineYO, dwibedi2021little}.
Leveraging nearest sample to produce pro views of sample mining is also proved effective in machine translation \cite{xin2021ANNMT} and language models \cite{Khandelwal2020GeneralizationTM}


\section{Conclusion}
In this paper, we propose a model-agnostic integrated cross-channel (\methodname~) approach, semantically connecting users and items for the matching stage of a typical industrial recommender system by maximally leveraging the inherent multi-channel mutual information.
Specifically, \methodname~ models correlation across user-item (U2I), user-user (U2U), and item-item (I2I) channels via intra and inter cross-channel contrastive modules.
\methodname~ naturally aligns users and items with semantic similarity and distinguishes them otherwise in each channel.
Extensive experiments show that our \methodname~ helps several popular retrieval models boost performance on four real-world benchmarks.
By deploying on industrial Tencent News platform with millions of users and conducting online experiments, we confirm the scalability and flexibility of the proposed method.

\bibliographystyle{ACM-Reference-Format}
\bibliography{main.bib}

\appendix
\section{Dataset and Metric}
\label{sec:datasetmetric}
We use four datasets as below:
\begin{itemize}
\item {\verb|Amazon Books|(\cite{2016upsanddowns})}: This dataset contains product reviews and metadata from Amazon, including 142.8 million reviews product metadata and links.
\item {\verb|Steam|}: This dataset contains more than 40k games from the steam shop with detailed data, including reviews and information about which games were bundled together.
\item {\verb|Taobao|\cite{10.1145/3219819.3219826}}: This dataset contains user behaviors recorded by Taobao recommendation system, consisting of users' clicks, item ID, item category, and timestamp. 
\item {\verb|Movielens-1M|\cite{Harper2015TheMD}}: One of the currently released MovieLens datasets, which contains 1,000,209 movie ratings from 6,040 users across 3,900 movies. 
\end{itemize}

\begin{table}[h]
\caption{Statistics of the Datasets.}
\centering
\resizebox{\columnwidth}{!}{%
\begin{tabular}{l cc cc c}
\toprule
Dataset & users & items & interactions  \\
    \midrule
    Amazon Books     & 459,133   & 313,966 & 8,898,041  \\
    Steam            & 2,567,538 & 15,474  & 7,793,069  \\ 
    Taobao          & 976,779 & 1,708,530 & 85,384,110 \\
    MovieLens-1M     & 6,040.    & 3,416   & 999,611    \\
    \bottomrule
\end{tabular}%
}
    \label{tab:staData}
\end{table}

The details of our evaluation metrics are as below:
\begin{itemize}
\item {\verb|Recall|}: Number of corrected recommended items divided by the total number of all recommended items.
\begin{equation}
    Recall@N = \frac{1}{|U|} \sum\limits_{u \in U} \frac{|\hat{I}_{u,N} \cap I_{u}|}{|I_u|}
\end{equation}
where ${\hat{I}_{u,N}}$ denotes the set of top-N recommended items for user u and ${I_u}$ is the set of testing items for user u.

\item {\verb|Normalized Discounted Cumulative Gain|}(NDCG): NDCG measures the percentage of correct recommended items, considering the positions of correct recommended items.
\begin{equation}
    DCG@N = \frac{1}{|U|} \sum\limits_{u\in U} \sum\limits_{r\in R} \frac{\delta_{N}(r)}{log_{2}(i_{r} + 1)},
\end{equation}
\begin{equation}
    NDCG@N = \frac{DCG@N}{IDCG@N}
\end{equation}
where G denotes the ground-truth list. ${i_{r}}$ is the index of r in R. ${\delta_{N}(\cdot)}$ is an indicator function which returns 1 if item r is in top-N recommendation, otherwise 0. IDCG is the DCG of ideal ground-truth list which refers to the descending ranking of ground-truth list in terms of predicted scores. 
\item {\verb|Hit Rate|(HR)}: This measures the percentage of at least one item is correctly recommended to and interacted by corresponding user.

\end{itemize}
\end{document}